\documentclass[aps,twocolumn,showpacs,preprintnumbers,prl]{revtex4}
\usepackage{graphicx}

\newcommand{\R}{{\bf r}}

\newcommand{\UP}{n_{\uparrow}}
\newcommand{\DN}{n_{\downarrow}}
\newcommand{\TUP}{\tau_{\uparrow}}
\newcommand{\TDN}{\tau_{\downarrow}}

\newcommand{\be}{\begin{equation}}
\newcommand{\ee}{\end{equation}}
\newcommand{\bea}{\begin{eqnarray}}
\newcommand{\eea}{\end{eqnarray}}
\newcommand{\bean}{\begin{eqnarray*}}
\newcommand{\eean}{\end{eqnarray*}}

\begin{document}

\title{Workhorse Semilocal Density Functional for Condensed Matter Physics\\ 
and Quantum Chemistry}
\author{John P. Perdew$^1$, Adrienn Ruzsinszky$^1$, G\'{a}bor I. Csonka$^2$, 
Lucian A. Constantin$^1$, and Jianwei Sun$^1$}
\affiliation{
$^1$Department of Physics and
Quantum Theory Group, Tulane University, New Orleans, LA 70118 USA\\
$^2$Department of Inorganic and Analytical Chemistry,
Budapest University of Technology and Economics,
H-1521 Budapest, Hungary}

\date{\today}

\begin{abstract}
Semilocal density functionals for the exchange-correlation energy are needed 
for large electronic systems. The Tao-Perdew-Staroverov-Scuseria (TPSS) 
meta-generalized gradient approximation (meta-GGA) is semilocal and usefully 
accurate, but predicts too-long lattice constants.  Recent "GGA's for solids" 
yield good lattice constants but poor atomization energies of molecules. We 
show that the construction principle for one of them (restoring the density 
gradient expansion for exchange over a wide range of densities) can be used to 
construct a "revised TPSS" meta-GGA with accurate lattice constants, surface 
energies, and atomization energies for ordinary matter.
\end{abstract}


\maketitle

Kohn-Sham theory \cite{a1} is the method of choice to describe large 
many-electron systems in condensed matter physics (since the 1970’s) and 
quantum chemistry (since the 1990’s).   In principle, this theory delivers the 
exact ground-state spin densities $\UP(\R)$, $\DN(\R)$ and energy $E$ for 
$N$ electrons in 
external potential $v(\R)$, via solution of $N$ selfconsistent one-electron 
Schroedinger equations. In practice, simple and reasonably accurate 
approximations to the density functional for the exchange-correlation energy 
are needed.  Semilocal approximations (e.g., Refs. \cite{a2,a3,a4}) 
% of the form
%
\begin{equation}
E_{xc}[\UP,\DN]=\int d^3r \; 
n\epsilon_{xc}(\UP,\DN,\nabla\UP,\nabla\DN,\TUP,\TDN)
\label{e1}
\end{equation}
%
%where $n=\UP+\DN$ is the electron density and 
%$\tau_{\sigma}=\sum|\nabla\psi_{i\sigma}|^2/2$ is the positive 
%kinetic energy density of 
%the occupied orbitals of spin $\sigma$ 
require only a single integral over 
real space 
and so are practical even for large molecules or unit cells.  
In Eq. (\ref{e1}), $n=\UP+\DN$ is the electron density and  
$\tau_{\sigma}=\sum_i|\nabla \psi_{i\sigma}|^2/2$
is the positive kinetic energy density; all 
equations are in atomic units.
Semilocal approximations often work because of proper accuracy for a 
slowly-varying density, or because
of justified error cancellation between exchange and correlation \cite{a5} (requiring a 
short-ranged xc hole).
They can be 
reasonably accurate for the near-equilibrium and compressed ground-state 
properties of "ordinary" matter, where neither strong correlation nor 
long-range van der Waals interaction are important.  They can also serve as a
base for the computationally more-expensive fully nonlocal approximations 
needed to describe strongly-correlated systems \cite{a5} and soft matter 
\cite{a6}. 

Semi-local functionals should be exact for the uniform electron gas, 
and should satisfy the spin- and coordinate-scaling properties of the exchange 
term $E_x$. The earliest one, the local spin density approximation (LSDA) 
\cite{a1,a2}, 
uses only the ingredients $\UP$
, $\DN$ and predicts reasonable but too-short lattice 
constants for solids, good surface energies for simple metals (but with 
substantial error cancellation between exchange and correlation), and 
molecular atomization energies that are unacceptably high.  The nonempirical 
Perdew-Burke-Ernzerhof  (PBE) generalized gradient approximation (GGA) \cite{a3} adds the ingredients $\nabla\UP$, 
$\nabla\DN$, and uses 
them 
to 
recover the gradient expansion for the correlation energy $E_c$ of a 
slowly-varying 
density, to make the correlation energy scale properly to a constant in the 
high-density  limit, and to satisfy other constraints.  The PBE GGA predicts 
reasonable but too-long lattice constants, surface energies that are better 
than LSDA for exchange alone and correlation alone but worse for their sum, 
and improved atomization energies.  The nonempirical TPSS meta-GGA \cite{a4} 
adds 
the ingredients $\TUP$, $\TDN$, and uses them to recover the fourth-order 
gradient expansion 
for exchange in the slowly-varying
limit, to make the functional exact for
the energy (but not for the potential) of all
one-electron ions, to make the exchange potential
finite at the nucleus, etc. 
TPSS predicts lattice constants that are only a little shorter 
than those of the PBE GGA, good surface energies, and very good atomization energies 
\cite{a4,a7}.  The bond lengths of stiff molecules 
are accurate \cite{a7} in TPSS.
A meta-GGA fitted to molecular data is M06-L \cite{ZT06}.

Meta-GGA is not computationally much more expensive than LSDA or GGA, 
once a selfconsistent program (e.g., Refs. \cite{a7} and \cite{a8}) has been 
written. 
For 
molecules containing 
transition-metal atoms, TPSS is only 30\% slower \cite{a8} than the PBE GGA. By 
respecting 
the paradigms of both condensed matter physics and quantum chemistry, the TPSS 
meta-GGA was intended to be a workhorse semilocal functional for both, and in 
particular for molecules bonded to or reacting on solid surfaces. Perhaps due 
to its lattice-constant errors, TPSS has not been so widely adopted. Because 
of the sensitivity of many solid state properties (magnetism, 
ferroelectricity \cite{WVK}, bulk modulus, etc.) to lattice constant, recent 
years have 
seen instead the emergence of "GGA's for solids" (e.g., the Armiento-Mattsson 2005
(AM05) GGA \cite{a9} and the modified PBE GGA for solids, 
PBEsol 
\cite{a10}) which typically predict good lattice constants and surface 
energies, 
but 
rather poor atomization energies.

The construction principle for the PBEsol GGA for solids \cite{a10} was to 
restore the second-order gradient expansion for exchange \emph{over a wide 
range of 
densities}.  Here we will show that this principle can be imposed to make a 
revised TPSS (revTPSS) meta-GGA that preserves all the correct constraints of 
TPSS, keeps its good surface and atomization energies, but yields lattice 
constants as good as those of the GGA's for solids.  We hope that revTPSS can 
become the workhorse functional that TPSS was intended to be. 

We begin with the semilocal exchange energy of a spin-unpolarized 
density \cite{a4}:
\begin{equation}
E^{sl}_x[n]=\int d^3r \; n\epsilon^{unif}_x(n)F_x(p,z).
\label{e2}
\end{equation}
Here $\epsilon^{unif}_x(n)=-3(3\pi^2n)^{1/3}/4\pi$ is the exchange energy per 
electron 
of a uniform gas of density $n$, $p=s^2$ is the square of 
the reduced density gradient $s=|\nabla n|/[2(3\pi^2n)^{1/3}n]$, and 
$z=\tau^W/\tau$ where $\tau^W=|\nabla n|^2/8n$ is 
the von 
Weizs\"{a}cker kinetic 
energy density and $\tau=\TUP + \TDN$.  The exchange enhancement factor 
$F_x$ is $1$ in LSDA, and 
otherwise $1+\kappa-\kappa/(1+x/\kappa)$, where $\kappa=0.804$. For a 
slowly-varying 
density, 
$x$ 
is small and of order $\nabla^2$, 
making $F_x\approx 1+x$. In GGA, $x=\mu p$, where $\mu=0.21951$ in the PBE GGA and 
$10/81=0.12346$ in the PBEsol GGA.  
In meta-GGA, $x$ 
depends upon $z$ as well as $p$, and only its slowly-varying asymptote is 
$(10/81)p$, 
but the large-$p$ asymptote of $F_x$, $1+\kappa-\kappa^2/\mu p$, is 
independent of 
$z$.  
As in Ref. \cite{a4}, we 
introduce $\alpha=(\tau-\tau^W)/\tau^{unif}=(5p/3)(z^{-1}-1)$, where 
$\tau^{unif}=n(3/10)(3\pi^2n)^{2/3}$ 
is the orbital kinetic energy density of the uniform gas.  
Any one- or two-electron density has $z=1$ or $\alpha=0$, while a 
slowly-varying density has  
small $z\approx 5p/3$ and $\alpha\approx 1$. 

In the TPSS meta-GGA, $x$ of the previous paragraph is given by Eq. (10) 
of Ref. \cite{a4}. For $\alpha\approx 1$ we can make the meta-GGA $F_x$ more 
like that of the PBEsol GGA through 
two changes : (1)  Change a term in $x$ from $cz^2p/(1+z^2)^2$ to 
$cz^3p/(1+z^2)^2$, which 
shifts this term (whose 
coefficient $c$ is much larger than a typical gradient coefficient) from 
$6^{th}$ to $8^{th}$ order in the gradient expansion.  All TPSS 
exchange constraints 
remain 
satisfied, 
without any change in the coefficients $c$, $e$, and $\mu$. $F_x$  is 
unchanged for $\alpha=0$,  
and at 
large $s$ for all $\alpha$, but is reduced at smaller $s$ for $\alpha=1$. The 
energy is raised more for a 
molecule (which has more regions of small $s$) than for the component atoms. 
%and 
%the mean meta-GGA error of the atomization energy is shifted from a small 
%positive to a small negative value.  
(2) Now change $\mu$ from its TPSS (and PBE) 
value $0.21951$ toward its PBEsol value $10/81$, letting $c$ and $e$ adjust 
accordingly to satisfy all TPSS constraints.  Fig. 1 shows a good emulation of 
the PBEsol GGA by $\mu=0.14$, $c=2.35204$ and $e=2.1677$, over the range of 
physical 
importance $0<s<3$, and especially 
for $s<1$ where the second-order gradient expansion for exchange is valid 
\cite{a10,a17}.  
Reducing  $\mu$ reduces $F_x$ at large $s$, raising the energy more for the 
component 
atoms 
(which have more regions of large $s$)  than for the molecule.
The net effect of these two changes is
to decrease atomization energies slightly, on average, and to increase surface 
energies.
% and pushing the 
%mean error of the atomization energies back toward the TPSS value.  In this 
%way, we improve meta-GGA lattice constants without worsening meta-GGA 
%atomization energies.
%
%%%%%%%%%%%%%%%%%%%%%%%%%%%%%%%%%%%%%%%%%%%%%%%%%%%%%
\begin{figure}
\includegraphics[width=\columnwidth]{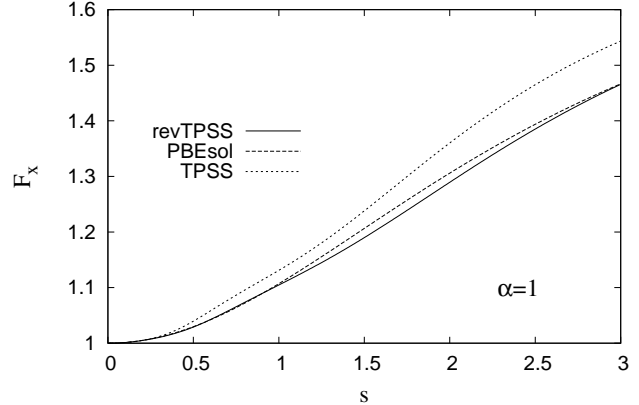}
\caption{Exchange enhancement factor vs. reduced density gradient for the 
PBEsol GGA
and for two meta-GGA’s at $\alpha=1$. The slowly-varying limit is 
$\alpha\approx 1$ and $s\approx 0$. By construction, 
revTPSS is closer to the PBEsol GGA than TPSS is.}
\label{f1}
\end{figure}
%%%%%%%%%%%%%%%%%%%%%%%%%%%%%%%%%%%%%%%%%%%%%%%%%%%%%%%

While the exact exchange energy is unique, the exact exchange energy 
density is not.  The conventional choice based on the Fock integral of the 
Kohn-Sham orbitals is just the $\lambda=1$ member of a one-parameter 
($0.5\leq\lambda\leq 1$) family of exact 
exchange energy densities, based upon a simple coordinate transformation 
\cite{a11}, 
which all agree for uniform densities but not for nonuniform ones. All have 
the same system-averaged exchange hole \cite{TSP}.  Figure 2 for the hydrogen 
atom shows 
that the choice $\lambda=0.893$ closely matches the revTPSS energy density of 
Eq. \ref{e2}.  Thus 
revTPSS has two reference systems in which it reproduces an exact 
exchange-correlation energy density: the uniform gas (a paradigm for condensed 
matter) and the one-electron atom or ion (a paradigm for quantum chemistry). 
(The same statement is true of TPSS \cite{a12}.)  Kohn and Mattsson \cite{a13} 
proposed 
supplementing the uniform gas reference system by one in which the density 
decays evanescently, but their second reference system was the Airy gas and 
not the hydrogen atom.  AM05 \cite{a9} exchange was constructed in part by 
fitting 
the conventional exchange energy density of the Airy gas \cite{Vi2}. Knowing 
the exact 
exchange energy density in the revTPSS or TPSS gauge may be useful for the 
construction of hyper-GGA's \cite{a5,a12}. 
%
%%%%%%%%%%%%%%%%%%%%%%%%%%%%%%%%%%%%%%%%%%%%%%%%%%%%%
\begin{figure}
\includegraphics[width=\columnwidth]{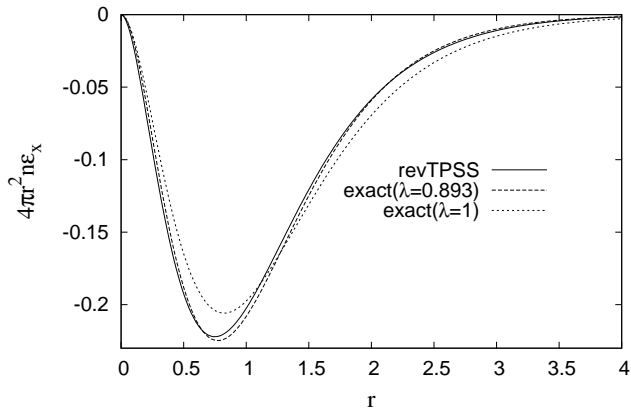}
\caption{Radial exchange energy density for the hydrogen atom. The revTPSS 
curve is from a spin-scaled Eq. (2).  Two $\lambda$-dependent exact exchange 
energy 
densities are 
also shown; the conventional one is $\lambda=1$.}
\label{f2}
\end{figure}
%%%%%%%%%%%%%%%%%%%%%%%%%%%%%%%%%%%%%%%%%%%%%%%%%%%%%%%

%The net effect of the changes we have made on TPSS exchange is to increase 
%atomization and surface energies. We can counter this effect through a 
%refinement of TPSS correlation.
Having improved TPSS exchange, we now refine TPSS correlation.
The PBE GGA and TPSS (through its ingredient 
$\epsilon^{PBE}_c(\UP,\DN,\nabla\UP,\nabla\DN)$) use a correlation gradient 
coefficient $\beta=0.066725$ derived in the high-density limit by Ma and 
Brueckner \cite{a14}.  
Langreth and Vosko \cite{a15} have derived a more correct value about 17\% 
bigger, 
but the difference comes from a long-range contribution to the gradient 
expansion of the correlation hole that would be cut off by our underlying 
real-space cutoff construction \cite{a16} of the PBE GGA.  We do not in any case need 
to 
restore the correct gradient coefficient for correlation, since for real 
densities the second-order gradient expansion for the correlation energy is 
never even close to being valid \cite{a10,a17}.  However, Hu and Langreth 
\cite{a18} have 
derived the density-dependence of the Ma-Brueckner $\beta$  beyond the random 
phase 
approximation, which \emph{is} relevant to our cutoff construction and which 
we have fitted roughly (Fig. 3) by
\begin{equation}
\beta(r_s)=0.066725(1+0.1r_s)/(1+0.1778r_s),
\label{e3}
\end{equation}
where $n=3/(4\pi r^3_s)$. Eq. (\ref{e3}) is designed so that, for 
$r_s\rightarrow\infty$, the second-order gradient terms for exchange and 
correlation cancel by innocuous assumption. 
%The meta-GGA atomization energies of molecules are 
%sensitive to $\beta(r_s)$, 
%especially when hydrogen atoms are involved: The free atom (and any 
%spin-polarized one-electron density) gets zero correlation energy for any 
%choice of $\beta(r_s)$, while the correlation energy of the hydrogen atom in 
%the molecule 
%must depend on $\beta(r_s)$. Our first guess, Eq. (3), turned out to be more 
%accurate for 
%atomization energies than is a better fit to the Hu-Langreth calculation.
The reduction of $\beta$ with $r_s$ increases atomization energies
slightly, on average, and decreases surface energies.
Aside from our use of Eq. (3), we keep the form of TPSS correlation unchanged, 
satisfying the TPSS constraints with
\begin{equation}
C(\zeta,0)=0.59+0.9269\zeta^2+0.6225\zeta^4+2.1540\zeta^6,
\label{e4}
\end{equation}
which replaces Eq. (13) of Ref. \cite{a4}.
%
%%%%%%%%%%%%%%%%%%%%%%%%%%%%%%%%%%%%%%%%%%%%%%%%%%%%%
\begin{figure}
\includegraphics[width=\columnwidth]{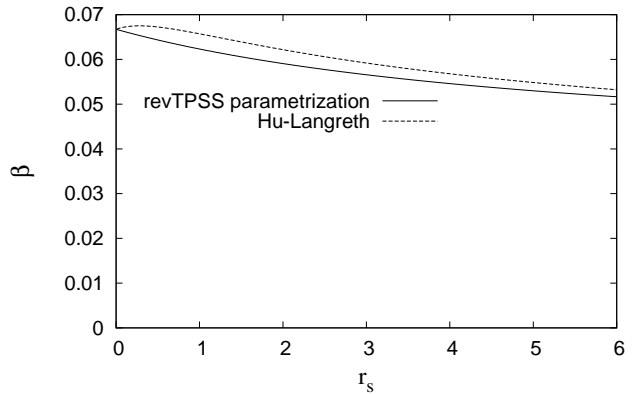}
\caption{Finite-range contribution to the gradient coefficient for 
correlation, as a function of density parameter $r_s$, from Ref. \cite{a18} 
and from Eq. (3).}
\label{f3}
\end{figure}
%%%%%%%%%%%%%%%%%%%%%%%%%%%%%%%%%%%%%%%%%%%%%%%%%%%%%%%

We turn now to the results, which are summarized briefly in Table 1 
(in terms of the mean error or ME and the mean absolute error or MAE, or 
their relative analogs MRE and MARE) and in 
full detail (along with figures for revTPSS $F_x$ and $F_{xc}$) in 
Ref. \cite{SI}.
Table I shows the error 
statistics of several density functionals for the lattice constants of 21 
solids , in comparison with experimental values corrected to a static lattice, 
calculated as in Ref. \cite{a19} using the BAND \cite{a20} code.  The 21 
solids include 11 
metals (Li, Na, Ca, Sr, Ba, Al, Pb, Cu, Rh, Pd, Ag) and 10 nonmetals (C 
diamond, Si, SiC, Ge, GaAs, NaCl, NaF, LiCl, LiF, MgO).  Our test set is the 
same as that of Ref. \cite{a19}, except that we have here omitted the three 
softest 
solids (K, Rb, Cs), for which the revTPSS lattice constants are about 
0.10-0.15 \AA $\;$ too long.  K, Rb, and Cs have bulk moduli (4 to 2 GPa 
\cite{a19}) close to those of the 
rare-gas solids Xe, Kr, and Ar \cite{AS} (and less than 1\% of that of 
diamond), and so could be classified as "soft 
matter" 
where the long-range van der Waals interaction between large ion cores can 
significantly shrink the lattice constant; we plan to investigate this in 
future work.  Table I shows that revTPSS performs about as well (and actually 
better) than the "GGA's for solids", the PBEsol GGA and the AM05 GGA.  Larger data sets 
\cite{a21} show error statistics similar to ours for the functionals that preceded 
revTPSS. After the lattice constants, Table I reports some cohesive energy results for solids.

Table I also shows the error statistics for the exchange- and 
exchange-correlation surface energies for jellium, computed as in Ref. 
\cite{a10}, 
in comparison to the exact exchange and nearly-exact revTPSS 
exchange-correlation 
values for bulk densities with  $r_s=2$, 3, 4, and 6.  Again, revTPSS performs 
well.  
Its very accurate surface exchange energies reflect \cite{a10,a17} its correct 
recovery of the gradient expansion for exchange.
\begin{table}
\caption{Error statistics for various density functionals.
(0.5292 \AA = 1 bohr.  
% $1.557\times 10^6 \rm{erg}/\rm{cm}^2 =1\rm{hartree}/\rm{bohr}^2$. 
1 kcal/mol=0.0434 eV =
0.00159 hartree.)  The exact jellium surface exchange-correlation energy is
still imprecisely known.  
We have taken it to be the revTPSS value, although it could instead
be the TPSS value as in Ref. \cite{a10}.
The non-revTPSS
lattice constants are from Ref. \cite{a19}.  Most non-revTPSS surface 
energies
and
atomization energies are from Ref. \cite{a10}. The non-revTPSS enthalpies of
formation are from Ref. \cite{a7}.}
\begin{ruledtabular}
\begin{tabular}{cccccccc}
\hline
 &LSDA&PBE&TPSS&AM05&PBEsol&revTPSS\\
\hline
\multicolumn{7}{|c|}{lattice constants (\AA) of 21 solids}\\
ME&-0.079&0.054&0.033&0.014&-0.010&0.011\\
MAE&0.079&0.065&0.047&0.039&0.038&0.036\\
%\hline
\multicolumn{7}{|c|}{cohesive energies of 9 non-transition}\\
\multicolumn{7}{|c|}{metals and 6 insulators (eV/atom)}\\
ME&0.32&-0.10&--&--&0.11&0.03\\ 
MAE&0.32&0.11&--&--&0.14&0.09\\
\multicolumn{7}{|c|}{jellium surface exchange energies 
(\%)}\\
\multicolumn{7}{|c|}{for $r_s=2$, 3, 4, 6}\\
%MRE&45.6&-20.9&-12.1&30.0&2.9&-1.0\\ 
%MARE&45.6&20.9&12.1&30.0&2.9&2.2\\
MRE&45.8&-20.9&-11.9&28.8&2.9&-1.0\\ 
MARE&45.8&20.9&11.9&28.8&2.9&2.2\\
%\hline
\multicolumn{7}{|c|}{jellium surface exchange-correlation energies 
(\%)}\\ 
\multicolumn{7}{|c|}{for $r_s=2$, 3, 4, 6}\\
%MRE&1.8&4.5&0.0&-2.7&-0.6&-1.1\\
%MARE&1.8&4.5&0.0&1.2&0.7&1.1\\
MRE&-2.9&-5.6&-0.8&0.6&-0.4&0.0\\
MARE&2.9&5.6&0.9&0.9&1.2&0.0\\
%\hline
\multicolumn{7}{|c|}{atomization energies (kcal/mol)}\\
\multicolumn{7}{|c|}{of the 6 AE6 molecules (6-311+G(3df,2p))}\\
ME&77.4&12.4&4.1&38.7&35.9&3.3\\
MAE&77.4&15.5&5.9&38.7&35.9&5.9\\
%\hline
\multicolumn{7}{|c|}{enthalpies of formation (kcal/mol)}\\
\multicolumn{7}{|c|}{of the 223 G3 molecules (6-311+G(3df,2p))}\\
ME&-121.9&-21.7&-5.1&--&--&-3.6\\
MAE&121.9&22.2&5.7&--&--&4.8\\
\hline
\end{tabular}
\end{ruledtabular}
\label{table2}
\end{table}

In Table I, we present the error statistics for the six atomization 
energies ( SiH$_4$, SiO, S$_2$, C$_3$H$_4$, C$_2$H$_2$O$_2$, C$_4$H$_4$ ) of 
the small representative AE6 \cite{a22} set 
and the 223 enthalpies of formation of the G3 \cite{a23} set, computed 
selfconsistently using a modified Gaussian code \cite{a24}. Note that, by 
construction, 
the 
error of the enthalpy of formation is nearly equal and opposite to that of the 
atomization energy.  The revTPSS values are good, and even a little better on 
average than the TPSS values. For the 47
G3 pure hydrocarbons, the MAE drops from 5.9 (TPSS) to 3.1 (revTPSS) kcal/mol.

Finally, as a check on hydrogen bonds, we have applied revTPSS to the 
W6 set \cite{a25} of dissociation energies of six small water clusters (four 
dimers, 
two trimers).  The revTPSS error statistics (ME=-1.0 kcal/mol, MAE=1.0 
kcal/mol) are only slightly worse than those of TPSS (ME=-0.9 kcal/mol, 
MAE=0.9 kcal/mol), but not as good as those of the PBE GGA (ME=-0.0 kcal/mol, MAE=0.3 
kcal/mol).  In the original TPSS, $\mu$ was set to the PBE GGA 0.21951 out of 
concern 
for the hydrogen bonds. Note that the inclusion of long-range van der Waals 
interaction could at least reduce the revTPSS error statistics for the W6 set.

Stereoelectronic effects on the energies of hydrocarbons are actually 
better described \cite{a26} by the PBEsol GGA than by the PBE GGA or TPSS.  We plan to test 
revTPSS 
for these problems.

In summary, we have shown that the PBEsol idea \cite{a10}, restoring the 
second-order gradient expansion for exchange over a wide range of densities, 
can be applied to the TPSS meta-GGA \cite{a4}, leading to a revised version 
(revTPSS) with good lattice constants, surface energies and atomization 
energies. revTPSS could well become a workhorse semilocal density functional 
for the ordinary matter of condensed matter physics and quantum chemistry, as 
well as a base for the construction of fully nonlocal approximate functionals.

We acknowledge support from the NSF 
under Grant DMR-0501588, and from MTA-NSF(98).

\end{document}